\documentclass[twocolumn,prd,showpacs]{revtex4-1}

\def\Lie{\pounds}
\def\dual{{}^\star\!}

\begin{document}
\title{Optimal Choices of Reference for a Quasi-local Energy: \\Spherically Symmetric Spacetimes}

\author{Ming-Fan Wu$^{1}$}\email{93222036@cc.ncu.edu.tw}
\author{Chiang-Mei Chen$^{1}$}\email{cmchen@phy.ncu.edu.tw}
\author{Jian-Liang Liu$^{1}$}\email{liujl@phy.ncu.edu.tw}
\author{James M. Nester$^{1,2}$}\email{nester@phy.ncu.edu.tw}

\affiliation{$^1$Department of Physics $\&$ Center for Mathematics and Theoretical Physics, National Central University, Chungli 320, Taiwan}
\affiliation{$^2$Graduate Institute of Astronomy, National Central University, Chungli 320, Taiwan}

\date{\today}



\begin{abstract}
For a given timelike displacement vector the covariant Hamiltonian quasi-local energy expression requires a proper choice of reference spacetime. We propose a program for determining the reference by  embedding a neighborhood of the two-sphere boundary in the dynamic spacetime into a Minkowski reference, so that the two sphere is embedded isometrically, and then extremizing the energy to determine the embedding variables. Applying this idea to Schwarzschild spacetime, we found that for each given future timelike displacement vector our program gives a unique energy value. The static observer
measures the maximal energy.
Applied to the Friedmann-Lema\^{i}tre-Robertson-Walker spacetime, we find that
the maximum energy value 
is nonnegative;
the associated displacement vector is the unit dual mean curvature vector, and 
the expansion of the two-sphere boundary matches 
that of its reference image.
 For these spherically symmetric cases the reference determined by our program is
 equivalent to isometrically matching the geometry at the two-sphere boundary
 and taking the displacement vector to be orthogonal to the spacelike constant coordinate time hypersurface,
 like the timelike Killing vector of the Minkowski reference.
\end{abstract}

\maketitle

\section{Introduction}

It is well-known that how to define gravitational energy is still an outstanding problem.
Since Einstein first published his general theory of relativity investigators have put much
effort into this problem, and some significant progress has been made. Many proposed expressions
for a gravitational energy-momentum density~\cite{Papapetrou:1948jw, Mol58, wein, tra, lan,
Bergmann:1953jz, Goldberg:1958zz}, but, caught by the equivalence principle, see,
e.g., \S 20.2 in~\cite{MTW}, they are all pseudotensors rather than tensors.
The modern concept, introduced by Penrose~\cite{Penrose:1982wp}, is that properly energy-momentum
is quasi-local, being associated with a closed surface bounding a region (for a comprehensive
review see~\cite{Szabados:2004vb}).
Various ideas about how to define a quasi-local energy have been proposed~\cite{Hawking:1968qt, Penrose:1982wp, Katz:1985, Katz:1990, Katz:2006uw, Jezierski:1990vu, Dougan:1991zz, Bergqvist:1992b, Nester:1994zn, Tung:1995cj, Robinson:1996, Hayward:1993ph, Brown:1992br, Lau:1993yv}. Not surprisingly the different definitions generally give different results~\cite{Bergqvist:1992a}. In this work we consider one
based on 
the covariant Hamiltonian formalism~\cite{Chen:1994qg, Chen:1998aw, Chang:1998wj, Chen:2000xw, Chen:2005hwa}. In that formalism one needs to choose a displacement vector and a reference spacetime to determine the energy;
%
although there was a little exploration of the issue in~\cite{Chen:1998aw}, no specific algorithm for choosing them was prescribed.

It is generally accepted that the gravitational energy for an asymptotically flat system should be
non-negative~\cite{Schoen:1979zz, Witten:1981mf}. Motivated by this property, it is reasonable to
presume that a suitable reference should be the one which minimizes the energy. The results of our
earlier explorations of this proposal (along with an alternative analytic approach)
are reported in~\cite{Liuthes}. A brief letter summarizing our later investigations has already
appeared~\cite{Chen:2009zd}. The present work is intended to be the first in a series giving a
detailed comprehensive exposition of our ideas. As in~\cite{Chen:2009zd}, in this paper we will
embed a two-sphere boundary $S$ and its neighborhood in the dynamic spacetime at a constant time and
radius into the Minkowski reference---with the 2-geometry embedded isometrically---and then extremize
the energy to fix the embedding, and thereby determine the reference. We use the Schwarzschild spacetime
to test our program. A number of investigators have used different quasi-local energy definitions to
calculate the energy of Schwarzschild spacetime,
many of them obtained the same results outside the horizon. However, there is no such consensus inside the horizon~\cite{Chen:2009zd, Booth:1998eh, Lundgren:2006fu, Yu:2008ij, Blau:2007wj}.
%
We should mention in particular that in~\cite{Yu:2008ij} the authors, using the Brown-York expression, not only get exactly the same results for radial geodesic observers as we do but also have the same physical interpretations about the observer dependence and the negative energy result. We show here that when applying this energy-extremization program to Schwarzschild spacetime for each future timelike displacement vector there is a uniquely corresponding energy but an equivalence class of the references. The reference is uniquely determined if we also impose the normalization condition of the displacement vector in the reference spacetime.

Our program allows us to calculate the energy measured by any physical observer. For measuring energy inside the black hole horizon, we first show the results measured by a radial geodesic observer. Then we examine several physical observers, we find they all get reasonable but different results inside the horizon. If we further vary the energy with respect to the displacement vector, then we get the maximum energy, which turns out to be that measured by the static observer. To further support our program we then consider a slightly different approach. We assume that the displacement vector is the timelike Killing vector in the Minkowski reference, and then do the extremization. This slightly different program  produces the same maximum energy as that produced by the original program. Although the results seem very reasonable, one might still wonder what does this energy-extremization program really mean geometrically.
 We show 
 that it is equivalent to matching the geometry near the two-sphere boundary, i.e., the reference is 4D isometric with the dynamic geometry on the boundary, and it makes the  displacement vector $N$ the unit normal of the constant coordinate time hypersurface, just like the timelike Killing vector of the Minkowski reference. This fact gives us more confidence in the program.

Applying this program to dynamic spherically symmetric spacetimes, one of the most interesting cases is surely homogeneous and isotropic cosmology, i.e., the Friedmann-Lema\^{i}tre-Robertson-Walker (FLRW) spacetimes. Many researchers have used various pseudotensor expressions both in general relativity and teleparallel gravity to calculate the energy of the FLRW spacetimes (see, e.g.,~\cite{Sharif:2008gq} and the references therein).
%
 Afshar~\cite{Afshar:2009pi} calculated the quasi-local energy of FLRW spacetime by using Brown-York's expression and got interesting results.
In~\cite{Nester:2008xd} the energy value determined for homogeneous cosmologies by the favored covariant Hamiltonian boundary term with a
\emph{homogeneous reference} was found to be
%
zero for all Bianchi class A models and negative for all class B models.
 In this paper we are going to show that the energy-extremization program likewise works well in dynamic spherically symmetric spacetimes and we give some discussion of  our results.

In the following section we briefly review the covariant Hamiltonian formulation. Then we apply our program to Schwarzschild spacetime in general coordinate systems in section 3 and to FLRW cosmology in section 4; the meaning of the energy extremization program results is discussed at the end of each of these sections.  
The conclusion includes a summary of the work.  

\section{Hamiltonian Formulation}
In this section we briefly review the covariant Hamiltonian formalism as developed by our research group~\cite{Chen:1994qg, Chen:1998aw, Chang:1998wj, Chen:2000xw, Chen:2005hwa} (for some additional developments along similar lines see~\cite{Anco:2001gk, Anco:2001gm}) and then apply it to Einstein's gravity theory: general relativity (GR).

A {\em first order Lagrangian 4-form} for a $k$-form field $\varphi$ has the form
\begin{equation} \label{1ordL}
{\cal L} = d \varphi \wedge p - \Lambda(\varphi,p).
\end{equation}
 Its variation
\begin{equation} \label{varL}
\delta {\cal L} = d(\delta \varphi \wedge p) + \delta \varphi \wedge \frac{\delta {\cal L}}{\delta \varphi} + \frac{\delta {\cal L}}{\delta p} \wedge \delta p,
\end{equation}
defines {\em a pair of first order} Euler-Lagrange expressions, which are explicitly given by
\begin{equation} \label{1ordfe}
\frac{\delta {\cal L}}{\delta p} := d \varphi - \partial_p \Lambda, \qquad \frac{\delta {\cal L}}{\delta \varphi} := - \varsigma d p - \partial_\varphi \Lambda,
\end{equation}
where $\varsigma := (-1)^k$.  The integral of ${\cal L}$ associates an action with any spacetime region.  The variation of this action is given by the integral of $\delta{\cal L}$. The total differential term in~(\ref{varL}) then leads to an integral over the boundary of the region. Hamilton's principle---that the action should be extreme with $\delta\varphi$ vanishing on the boundary---yields the field equations: the vanishing of the Euler-Lagrange expressions~(\ref{1ordfe}).

The action should not depend on the particular way points are labeled.  Thus it should be invariant under diffeomorphisms, including infinitesimal diffeomorphisms---which correspond to a displacement along some vector field $N$. From a gauge theory perspective such displacements are regarded as  ``local translations''. Under a local translation quantities change according to the Lie derivative. Hence, for a diffeomorphism invariant action
the relation~(\ref{varL}) should be identically satisfied when the variation operator $\delta$ is replaced by the Lie derivative $\Lie_N$ ($\equiv di_N+i_Nd$ on the components of form fields):
\begin{equation}
d i_N {\cal L} \equiv \Lie_N{\cal L} \equiv  d(\Lie_N \varphi \wedge p) + \Lie_N \varphi \wedge \frac{\delta {\cal L}}{\delta \varphi} + \frac{\delta {\cal L}}{\delta p} \wedge \Lie_N p. \label{noethertrans}
\end{equation}
This simply means that ${\cal L}$ is a 4-form which depends on position only through the fields $\varphi, p$. (For this to be the case the set of fields in ${\cal L}$ necessarily includes dynamic geometric variables, which means gravity.)

From the identity~(\ref{noethertrans}) it directly follows that the 3-form
\begin{equation}\label{HN}
{\cal H}(N) := \Lie_N \varphi \wedge p - i_N {\cal L},
\end{equation}
satisfies the identity
\begin{equation} \label{IdH}
- d {\cal H}(N) \equiv  \Lie_N \varphi \wedge \frac{\delta {\cal L}}{\delta \varphi} + \frac{\delta {\cal L}}{\delta p} \wedge \Lie_N p,
\end{equation}
showing that it is a conserved ``current'' {\em on shell} (meaning: when the field equations are satisfied). Substituting the Lagrangian 4-form~(\ref{1ordL}) into the definition~(\ref{HN}) leads to the explicit expression $ {\cal H}(N) \equiv d(i_N \varphi \wedge p) + \varsigma i_N \varphi \wedge d p + \varsigma d \varphi \wedge i_N p + i_N \Lambda $, from which one can see that this conserved  {\ Noether translation current} can be written as a 3-form linear in the displacement vector plus a total differential:
\begin{equation}
{\cal H}(N) =: N^\mu {\cal H}_\mu + d {\cal B}(N). \label{H+dB}
\end{equation}
Compare the differential of this expression, $ d{\cal H}(N) \equiv dN^\mu\wedge {\cal H}_\mu + N^\mu d {\cal H}_\mu, $ with the identity~(\ref{IdH}); equating the $dN^\mu$ coefficient on both sides reveals that
\begin{equation}
{\cal H}_\mu \equiv - i_\mu \varphi \wedge \frac{\delta {\cal L}}{\delta \varphi} + \varsigma \frac{\delta {\cal L}}{\delta p} \wedge i_\mu p.
\end{equation}
This (Noether's second theorem type) identity is a necessary consequence of {\em local} diffeomorphism invariance (i.e., a symmetry for non-constant $N^\mu$). From this relation one can see that ${\cal H}_\mu$ vanishes on shell; hence the value of the conserved quantity associated with a local displacement $N$ and a spatial region $\Sigma$ is determined by a 2-surface integral over the boundary of the region:
\begin{equation}
E(N, \Sigma) := \int_\Sigma {\cal H}(N) = \oint_{\partial\Sigma} {\cal B}(N). \label{EN}
\end{equation}
The value is {\em quasi-local} (depending only on the values of the fields on the boundary).  For {\em any} choice of $N$ this expression defines a conserved quasi-local quantity.

One can adjust by hand the 2-form ${\cal B}(N)$ without affecting the above argument.  Of course this will modify the value of the quasi-local quantities.  However this freedom in choosing the boundary term is entirely under physical control, since ${\cal H}(N)$ is not merely the translational Noether current 3-form but is also the {\em Hamiltonian 3-form} which generates the changes (given by $\Lie_N$) in the dynamical quantities.  The boundary term in the variation of the Hamiltonian reflects the necessary boundary conditions implicit in the Hamiltonian.   Consequently changing ${\cal B}(N)$ will likewise modify the boundary conditions implicit in ${\cal H}(N)$. Thus different choices of boundary term are associated with different physical boundary conditions.

The above applies quite generally to any geometric (i.e., diffeomorphically invariant) gravity theory.  Here we confine our attention specifically to  Einstein's gravity theory, general relativity (GR).  GR can be formulated in several ways. For our purposes the most convenient is to take the {\em orthonormal coframe} $\vartheta^\mu = \vartheta^\mu{}_k dx^k$ and the {\em connection one-form} coefficients $\omega^\alpha{}_\beta = \Gamma^{\alpha}{}_{\beta k} dx^k$ as our geometric potentials.  Moreover we take the connection to be {\em a priori} metric compatible:
$Dg_{\alpha\beta} := dg_{\alpha\beta} - \omega^\gamma{}_\alpha g_{\gamma\beta} - \omega^\gamma{}_\beta g_{\alpha\gamma} \equiv 0$. Restricted to orthonormal frames where the metric coefficients are constant, this condition reduces to the algebraic constraint $\omega^{\alpha\beta} \equiv \omega^{[\alpha\beta]}$.

We consider the vacuum (source free) case for simplicity (the inclusion of sources is straightforward).  GR can be obtained from the first order Lagrangian 4-form
\begin{equation}
{\cal L}_{\rm GR} = \Omega^{\alpha\beta} \wedge \rho_{\alpha\beta} + D\vartheta^\mu \wedge \tau_\mu - V^{\alpha\beta} \wedge ( \rho_{\alpha\beta} - \frac1{2\kappa} \eta_{\alpha\beta} ),
\end{equation}
where $\Omega^\alpha{}_\beta := d \omega^\alpha{}_\beta + \omega^\alpha{}_\gamma \wedge \omega^\gamma{}_\beta$ is the {\em curvature} 2-form,
$D\vartheta^\mu := d\vartheta^\mu + \omega^\mu{}_\nu \wedge \vartheta^\nu$ is the {\em torsion} 2-form,  and we have made use of the convenient
dual form basis $\eta^{\alpha\beta\dots} := \dual (\vartheta^\alpha \wedge \vartheta^\beta\wedge \cdots)$. The 2-forms $\Omega^{\alpha\beta}$, $V^{\alpha\beta}$ and $\rho_{\alpha\beta}$ are antisymmetric. We take $\kappa := 8\pi G/c^4 = 8 \pi$. Several possible boundary terms were identified, each associated with a distinct
type of boundary condition.  In~\cite{Chen:2005hwa} a ``preferred boundary term'' (it has a certain covariant property, directly gives the Bondi energy flux, and
has a positive total energy proof) for GR was identified:
\begin{equation} \label{expB}
B(N) = \frac{1}{16\pi} (\Delta\Gamma^{\alpha}{}_{\beta} \wedge \iota_{N} \eta_{\alpha}{}^{\beta} + \bar D_{\beta} N^\alpha \Delta \eta_{\alpha}{}^\beta),
\end{equation}
where $\Delta$ indicates the difference between the dynamic and reference values, and $\bar D_{\beta}$ is the covariant derivative using the reference connection. The reference values can be determined by pullback from
an embedding of a neighborhood of the boundary into a suitable reference space. Now we can use this expression to calculate the gravitational energy in general relativity.

\section{Static spacetime: Schwarzschild spacetime}
\subsection{Energy-Extremization Program}
Our objective is to find a principle for determining the displacement vector and reference spacetime so that we can calculate the quasi-local energy for gravitating systems. To achieve that we have proposed a program including isometric embedding and energy extremization. Now we're going to use the Schwarzschild metric to test this program.

The Schwarzschild metric in the standard spherical coordinate form is given by
\begin{equation}
ds^2 = - A dt^2 + A^{-1} dr^2 + r^2 d\Omega_2^2,
\end{equation}
where $A = 1 - 2m/r$ and $d\Omega_2^2 = d\theta^2 + \sin^2\theta d\phi^2$. However, there are several other reasonable coordinate choices for the Schwarzschild metric, such as Painlev\'e-Gullstrand, Eddington-Finkelstein, and Kruskal-Szekeres. In order to accommodate most well-known coordinates, we consider a more general version of the Schwarzschild metric via a coordinate transformation $t = t(u,v), r = r(u,v)$
\begin{eqnarray} \label{genMetric}
ds^2 &=& - ( A t^2_{u} - A^{-1} r^2_{u} ) du^2 + 2 ( A^{-1} r_{u} r_{v} - A t_{u} t_{v} ) dudv
\nonumber\\
&& + ( A^{-1} r^2_{v} - A t^2_{v} ) dv^2 + r^2 d\Omega_2^2.
\end{eqnarray}
Note that under this coordinate transformation we have assumed that the orientation is preserved so that the Jacobian is positive, i.e., $\sqrt{-\alpha} := t_u r_v - t_v r_u > 0$. Choose Minkowski spacetime as the reference:
\begin{equation}
d\bar{s}^2 = - dT^2 + dR^2 + R^2 d\Theta^2 + R^2 \sin^2\Theta d\Phi^2.
\end{equation}
The essential issue of the reference choice is the identification between the reference and physical spacetime coordinates. A legitimate approach for the spherically symmetric case is to assume $T = T(u,v),\ R = R(u,v),\ \Theta = \theta,\ \Phi = \varphi$ along with $R_0 := R(t_0, r_0) = r_0$; this symmetrically embeds a neighborhood of the two-sphere boundary $S$ at $(t_0, r_0)$ into the Minkowski reference such that the two-sphere boundary is embedded isometrically.

For any given future timelike displacement vector
\begin{equation}
N = N^u \partial_u + N^v \partial_v = N^t \partial_t + N^r \partial_r = N^T \partial_T + N^R \partial_R,\label{disp N}
\end{equation}
We expect that the second term of the expression~(\ref{expB}) wont't contribute because the spacetime is spherically symmetric. In particular, if we assume
\begin{eqnarray}
0 &=& \bar{D}_T N^R = \partial_T N^R + \bar{\Gamma}^R{}_{TT} N^T + \bar{\Gamma}^R{}_{RT} N^R = \partial_T N^R,
\nonumber\\
0 &=& \bar{D}_R N^T = \partial_R N^T + \bar{\Gamma}^T{}_{TR} N^T + \bar{\Gamma}^T{}_{RR} N^R = \partial_R N^T,
\nonumber\\
\label{0=DN}
\end{eqnarray}
where all the corresponding connection terms vanish, then so does the second term of~(\ref{expB}). Now from~(\ref{EN}, \ref{expB}) we get for the quasi-local energy associated with a sphere of radius $r$
\begin{eqnarray} \label{expE}
E(r) &=& \oint \frac{1}{16\pi} ( \Delta \Gamma^{\alpha}{}_{\beta} \wedge i_{N} \eta_{\alpha}{}^{\beta} + \bar D_{\beta} N^\alpha \Delta \eta_{\alpha}{}^\beta )
\nonumber \\
&=& \frac{r}2 \left( N^u B + N^v C \right) \sqrt{-\alpha}, \label{Eexp}
\end{eqnarray}
where
\begin{eqnarray}
B &=& X T_u + g^{vu} (R_u - 2 r_u) + g^{vv} (R_v - 2 r_v),
\nonumber\\
C &=& X T_v + g^{uu} (2 r_u - R_u) + g^{uv} (2 r_v - R_v),
\nonumber\\
X &=& (T_u R_v - T_v R_u)^{-1},
\end{eqnarray}
and the subscripts indicate the related partial differentiations. Note that the quasi-local energy is evaluated on the boundary, the two-sphere $S$; all the variables appearing in~(\ref{expE}) and in the following calculations are also evaluated on the two-sphere. Each choice of $\{T_u, T_v, R_u, R_v\}$, which we call the embedding variables, means a different embedding, hence a different reference. Now extremizing the energy with respect to the embedding variables we get the conditions
\begin{eqnarray}
\frac{\partial E}{\partial T_u} = 0 &\Rightarrow&  -X^2 T_v (N^u R_u + N^v R_v) = 0, \label{ETu}
\\
\frac{\partial E}{\partial T_v} = 0 &\Rightarrow& X^2 T_u (N^u R_u + N^v R_v) = 0, \label{ETv}
\\
\frac{\partial E}{\partial R_u} = 0 &\Rightarrow& X^2 T_v (N^u T_u + N^v T_v)
\nonumber\\  && + N^u g^{uv} - N^v g^{uu} = 0, \label{ERu}
\\
\frac{\partial E}{\partial R_v} = 0 &\Rightarrow& -X^2 T_u (N^u T_u + N^v T_v)
\nonumber\\  && + N^u g^{vv} - N^v g^{uv} = 0. \label{ERv}
\end{eqnarray}
Note that conditions~(\ref{ETu}, \ref{ETv}) are equivalent (since we do not want both $T_u$ and $T_v$ to vanish), so we only have three independent equations. Using the relations
\begin{equation}
g^{uu} = \alpha^{-1} g_{vv}, \quad g^{uv} = - \alpha^{-1} g_{uv}, \quad g^{vv} = \alpha^{-1} g_{uu},
\end{equation} with
$\alpha = - (t_u r_v - t_v r_u)^2$, then the conditions~(\ref{ETu}--\ref{ERv}) become
\begin{eqnarray}
N^u R_u + N^v R_v = N^R &=& 0,\quad \label{ETuv}
\\
X^2 T_v (N^u T_u \!+\! N^vT_v) \!-\! \alpha^{-1} (g_{uv} N^u \!+\! g_{vv} N^v) &=& 0,\quad \label{ERu2}
\\
X^2 T_u (N^u T_u \!+\! N^v T_v) \!-\! \alpha^{-1} (g_{uu} N^u \!+\! g_{uv} N^v) &=& 0.\quad \label{ERv2}
\end{eqnarray}
From $(\ref{ERv2})\times R_v -(\ref{ERu2})\times R_u$ we get
\begin{eqnarray}
X (N^u T_u + N^v T_v) + \alpha^{-1} \Bigl[ (g_{uv} N^u + g_{vv} N^v) R_u
\nonumber\\
- (g_{uu} N^u + g_{uv} N^v) R_v \Bigr] = 0. \label{ERuv}
\end{eqnarray}
Using condition~(\ref{ETuv}) we get
\begin{equation}
R_u = - \frac{N^v}{N^u} R_v, \label{Ru}
\end{equation}
and
\begin{eqnarray}
N^T &=& N^u T_u + N^v T_v
= \frac{N^u}{R_v} (T_u R_v - T_v R_u)
\nonumber\\
&=& \frac{N^u}{X R_v} \label{NT}
\end{eqnarray}
then $R_v$ can be found from (\ref{ERuv})
\begin{equation}
\frac{N^u}{R_v} - \alpha^{-1} \frac{R_v}{N^u} g(N,N) = 0 \quad \Rightarrow \quad R_v^2 = \frac{\alpha (N^u)^2}{g(N,N)}.
\end{equation}
Note that we require the displacement vector to be future timelike, i.e., $N^T > 0$ and $N^u > 0$, and the orientation to be preserved, i.e., the Jacobian is positive. From condition~(\ref{NT}) we know that $R_v$ should be positive, so
\begin{equation}
R_v = \sqrt{\frac{\alpha}{g(N,N)}} N^u, \quad R_u = - \sqrt{\frac{\alpha}{g(N,N)}} N^v. \label{RuRv}
\end{equation}
Now we can calculate the energy. Using the conditions~(\ref{NT}, \ref{RuRv}) we get
\begin{eqnarray}
&& N^u B + N^v C
\nonumber\\
&=& \frac{N^u}{R_v} + \alpha^{-1} \Bigl[ \sqrt{\alpha g(N,N)} + 2 (g_{uv} N^u r_u
\nonumber\\
&& - g_{uu} N^u r_v + g_{vv} N^v r_u - g_{uv} N^v r_v) \Bigr]
\nonumber\\
&=& 2 (t_u r_v - t_v r_u)^{-1} \left( \sqrt{-g(N,N)} - A N^t \right),
\end{eqnarray}
where we have used the metric in the calculation.  Choose $N$ to be unit timelike on the two-sphere:
\begin{eqnarray}
-1 &=& g(N,N)|_{S}
\nonumber\\
&=& g_{uu} (N^u)^2 + 2 g_{uv} N^u N^v + g_{vv} (N^v)^2, \label{unitN}
\end{eqnarray}
 then we get the quasi-local energy for any given future timelike displacement vector $N$:
\begin{equation}
E_{\rm ex}(N) = r \left[ 1 - A (N^u t_u + N^v t_v) \right] = r \left( 1 - A N^t \right),  \label{Energy}
\end{equation}
a result which is independent of any $u,v$ coordinate system choice. It is also independent of the two embedding variables $T_u, T_v$. Indeed, we cannot solve for all four variables, since we only have three independent conditions. So this program produces a unique energy but an equivalence class of references for any physical observer.

However, it is reasonable to impose the normalization condition of the displacement vector in the reference spacetime, i.e.,
\begin{equation}
-1 = \bar{g}(N,N) = -(N^T)^2 + (N^R)^2.
\end{equation}
which together with~(\ref{ETuv}) implies $N^T = 1$. Then from~(\ref{NT}) and~(\ref{RuRv}) we get
\begin{equation}
X = \frac{N^u}{R_v} = \frac{1}{\sqrt{-\alpha}}.
\end{equation}
Using this condition along with~(\ref{ERu2}, \ref{ERv2}) to solve for $T_u$ and $T_v$ we get
\begin{eqnarray}
T_u &=& A t_u N^t - A^{-1} r_u N^r,
\nonumber\\
T_v &=& A t_v N^t - A^{-1} r_v N^r.
\end{eqnarray}
In this way we can calculate for any given observer our Hamiltonian boundary term quasi-local energy of the Schwarzschild spacetime.

Now if we further vary the energy with respect to the displacement vectors, it should tell us which observer would measure the extreme energy. So from
\begin{equation}
g(N,N) = - A (N^t)^2 + A^{-1} (N^r)^2 = -1,
\end{equation}
we take
\begin{equation}
\sqrt{A} N^t = \cosh z, \quad \frac{1}{\sqrt{A}} N^r = \sinh z.
\end{equation}
Then we find
\begin{eqnarray}
\frac{\partial E}{\partial z} = 0 &\Rightarrow& \cosh z = 1,
\nonumber\\
&\Rightarrow& N^t = \frac{1}{\sqrt{A}}, \quad N^r = 0,
\nonumber\\
&\Rightarrow& E_{\rm max} = r \left( 1 - \sqrt{A} \right), \label{Emax}
\end{eqnarray}
and
\begin{equation}
\frac{\partial^2 E}{\partial z^2}\Big|_{\frac{\partial E}{\partial z} = 0} \leq 0.
\end{equation}
So among all physical observers the static observer, i.e., $N = \frac{1}{\sqrt{A}} \partial_t$,  measures the maximum energy~(\ref{Emax}).
This energy value
\begin{equation}
E_{\rm max}= r \left( 1 - \sqrt{1-2m/r} \right)\equiv \frac{2m}{1+\sqrt{1-2m/r}}\label{E-max}
\end{equation}
is a standard result~\cite{Brown:1992br, Chen:1994qg, Chen:1998aw, Liuthes, Chen:2009zd, Liu:2003bx, Wang:2008jy}. 

Now let us instead assume that the displacement vector is the timelike Killing vector in the Minkowski reference and then extremize the energy. From the previous sections we might already expect that the result should be also the value~(\ref{Emax}). Starting from our energy expression~(\ref{Eexp}) we assume
\begin{eqnarray}
N &=& \partial_T = N^u \partial_u + N^v \partial_v,
\nonumber\\
\Rightarrow \quad N^u &=& X R_v, \quad N^v = - X R_u, \label{NuNv}
\\
-1 &=& g(N,N)
\nonumber\\
&=& g_{uu} (N^u)^2 + 2 g_{uv} N^u N^v + g_{vv}(N^v)^2
\nonumber\\
&=& X^2 (g_{uu} R_v^2 - 2 g_{uv} R_u R_v + g_{vv} R_u^2)
\nonumber\\
\Rightarrow \quad X^2 &=& \frac{-1}{g_{uu} R_v^2 - 2 g_{uv} R_u R_v + g_{vv} R_u^2}. \label{X2}
\end{eqnarray}
Then the general energy expression~(\ref{Eexp}) becomes
\begin{eqnarray}
E &=& \frac{r}{2} X \Bigl[ 1 - \alpha^{-1} X^{-2} + 2 \alpha^{-1} (g_{uv} r_u R_v - g_{uu} r_v R_u
\nonumber\\
&& - g_{vv} r_u R_u + g_{uv} r_v R_u) \Bigr] \sqrt{-\alpha},
\end{eqnarray}
where (\ref{NuNv}, \ref{X2}) have been used. Now there are only two variables in our energy expression. We extremize the energy with respect to these two variables:
\begin{eqnarray}
\frac{\partial E}{\partial R_u} = 0 &\Rightarrow& (g_{vv} R_u - g_{uv} R_v) (1 + \alpha^{-1} X^{-2})
\nonumber\\
&& + 2 R_v (r_u R_v - r_v R_u) = 0, \label{ERuA}
\\
\frac{\partial E}{\partial R_v} = 0 &\Rightarrow& (g_{uu} R_v - g_{uv} R_u) (1 + \alpha^{-1} X^{-2})
\nonumber\\
&& - 2 R_u (r_u R_v - r_v R_u) = 0. \label{ERvA}
\end{eqnarray}
The combination $(\ref{ERuA}) \times R_u + (\ref{ERvA}) \times R_v$ implies
\begin{eqnarray}
0 &=& (1 \!+\! \alpha^{-1} X^{-2}) (g_{uu} R_v^2 \!-\! 2 g_{uv} R_u R_v \!+\! g_{vv} R_u^2)
\nonumber\\
&=& - X^{-2} (1 + \alpha^{-1} X^{-2})
\\
\Rightarrow \quad X &=& \sqrt{- \alpha^{-1}} = (t_u r_v - t_v r_u)^{-1}.
\end{eqnarray}
Now from~(\ref{ERuA}, \ref{ERvA}) we learn that
\begin{equation}
r_u R_v - r_v R_u = 0. \label{rR}
\end{equation}
Together with (\ref{X2}) this 
gives
\begin{eqnarray}
\alpha &=& -X^{-2} = g_{uu} R_v^2 - 2 g_{uv} R_u R_v + g_{vv} R_u^2
\nonumber\\
&=& \left( \frac{R_v}{r_v} \right)^2 (g_{uu} r_v^2 - 2 g_{uv} r_u r_v + g_{vv} r_u^2)
\nonumber\\
&=& \left( \frac{R_v}{r_v} \right)^2 A \alpha \nonumber
\end{eqnarray}
so we get
\begin{equation}
R_u^2 = \frac{r_u^2}{A}, \quad R_v^2 = \frac{r_v^2}{A}.
\end{equation}
Pick the plus sign, then we get the unique energy produced by this program:
\begin{equation}
E(\partial_T)_{\rm ex} = r \left( 1 - \sqrt{A} \right) \equiv E_{\rm max} \equiv E_{\rm ex}\left( \frac{\partial_t}{\sqrt{A}} \right) \label{ETmax}
\end{equation}
as expected.

\subsection{Quasi-local energy measured by other physical observers}
It is obvious that the energy value~(\ref{Emax}) is only valid outside the black hole horizon, since there is no static observer inside. Since our energy formula~(\ref{Energy}) can be applied to any physical observer, to measure the energy inside the black hole let us first examine the radial geodesic observers in the Schwarzschild spacetime. For an observer who falls initially with velocity $v_0$ from a constant $r = a$, there would be two different orbits available: (1) the crash orbit, where ingoing observers crash directly into the singularity at $r = 0$ and outgoing observers who shoot out to the turning point $r_\mathrm{max}$ and then return back to crash; (2) the crash/escape orbit, where ingoing observers crash and outgoing observers escape to infinity. We choose the unit tangent of the geodesic to be the suitable displacement vector, i.e.,
\begin{equation}
N^t = \frac{1}{1 - \frac{2 m}{r}} \frac{\sqrt{1 - \frac{2m}{a}}}{\sqrt{1 - v^2_0}},
\end{equation}
where $2m < a$ and $0 < r \leq r_\mathrm{max}$. The Hamiltonian boundary term energy value for this observer is
\begin{equation}
E = r \left( 1 - \frac{\sqrt{1 - \frac{2m}{a}}}{\sqrt{1 - v^2_0}} \right).
\end{equation}
This result also agrees with that found using the Brown-York quasi-local energy expression~\cite{Blau:2007wj, Yu:2008ij}. One can see that the energy decreases as the initial velocity $v_0$ increases. When the initial velocity $v_0$ is less, equal or greater than $\sqrt{2m/a}$, which is the escape velocity from the Newtonian point of view, the energy is positive, zero or negative, respectively.

The negative value for the energy may appear odd, but it can be explained physically. Just as was found in another instance~\cite{Nester:2008xd}, it
is correlated with the geometric property that the scalar curvature of the spacelike hypersurface orthogonal to the displacement vector is (unlike the usual cases) negative.

To see that let us consider a hypersurface orthogonal to the displacement vector
\begin{equation}
N = \frac{1}{1 - \frac{2 m}{r}} \frac{\sqrt{1 - \frac{2m}{a}}}{\sqrt{1 - v^2_0}} \partial_t + \left( \frac{1 - \frac{2m}{a}}{1 - v_0^2} - (1 - \frac{2m}{r}) \right)^{1/2} \partial_r.
\end{equation}
Then the induced metric on the hypersurface is
\begin{equation}
h_{\mu\nu} = g_{\mu\nu} + N_{\mu} N_{\nu},
\end{equation}
where $g_{\mu\nu}$ is the four dimensional spacetime metric. Using this induced metric one can compute the scalar curvature $R$ of this hypersurface,
which turns out to be
\begin{equation}
R = \frac{2}{r^2} \frac{\frac{2m}{a} - v_0^2}{1 - v_0^2}.
\end{equation}
One can see that the scalar curvature of this hypersurface becomes negative when the initial velocity is greater than the escape velocity, i.e.
\begin{equation}
v_0 > \sqrt{\frac{2m}{a}} \quad \Rightarrow \quad R < 0.
\end{equation}
Note that the ingoing geodesic observers can measure energy inside the black hole. This energy is proportional to the radial distance $r$, so it is a smooth function in the region $0 \leq r \leq r_\mathrm{max}$.

In addition to the static and radial geodesic observers there are other natural choices---notably the unit normal of the constant coordinate time hypersurface in various coordinate systems.

As an example consider the ADM form of (ingoing) {\em Eddington-Finkelstein} coordinates~\cite{Liuthes}:
\begin{equation}
ds^2 = - \frac{du^2}{1 + \frac{2m}{r}} + \left[ \sqrt{1 + \frac{2m}{r}} dv+\frac{2 m  du}{r \sqrt{1 + \frac{2m}{r}}}   \right]^2 + r^2 d\Omega_2^2,
\end{equation}
where
\begin{eqnarray}
&& du = dt + \frac{2m}{r}A^{-1} dr, \qquad dv = dr
\nonumber\\
\Rightarrow && t_u = 1, \quad t_v = - \frac{2m}{r} A^{-1}, \quad r_u = 0, \quad r_v = 1. \nonumber
\end{eqnarray}
For this choice the time-component of the coframe and the displacement vector are
\begin{eqnarray}
\vartheta^0 &=& \frac{du}{\sqrt{1 + \frac{2m}{r}}},
\nonumber\\
N &=& e_0 = \sqrt{1 + \frac{2m}{r}} \; \partial_u- \frac{2m}{r \sqrt{1 + \frac{2m}{r}}} \; \partial_v, \nonumber
\end{eqnarray}
consequently
\begin{equation}
N^t = N^u t_u + N^v t_v = A^{-1} \left( 1 + \frac{2m}{r} \right)^{-\frac12},
\end{equation}
and the energy~(\ref{Energy}) then has the value
\begin{eqnarray}
E_{\rm EF} &=& r \left( 1 - ({1 + {2m}/{r}})^{-1/2} \right)
\nonumber\\
&=& 2 m \left( 1 + {2m}/r + \sqrt{1+{2m}/r} \right)^{-1}.
\end{eqnarray}

For (ingoing) {\em Painlev\'e-Gullstrand\/} coordinates:
\begin{equation}
ds^2 = - du^2 + \left(dv+ \sqrt{\frac{2m}{r}} du  \right)^2 + r^2 d\Omega_2^2,
\end{equation}
where
\begin{equation}
du = dt + A^{-1} \sqrt{\frac{2m}{r}}dr, \qquad dv = dr.
\end{equation}
The time-component of the coframe and the displacement vector are
\begin{eqnarray}
&& \vartheta^0 = du, \qquad N = e_0 = \partial_u - \sqrt{\frac{2m}{r}} \partial_v,
\nonumber\\
\Rightarrow && t_u = 1, \quad t_v = -A^{-1} \sqrt{\frac{2m}{r}}, \quad r_u = 0, \quad r_v = 1. \nonumber
\end{eqnarray}
consequently
\begin{equation}
N^t = N^{\mu} t_{\mu} + N^{\nu} t_{\nu} = A^{-1},
\end{equation}
and the value of the energy~(\ref{Energy}) now works out to be
\begin{equation}
E_{\rm PG} = 0.
\end{equation}
For this special case the energy {\em vanishes} for all $r$ (note that the ADM {\em total energy} also vanishes for this form of the metric).

In {\em Kruskal-Szekeres} coordinates
\begin{eqnarray}
ds^2 &=& \frac{32m^3}{r} \; \mathrm{e}^{-\frac{r}{2m}} (- du^2 + dv^2) + r^2 d\Omega_2^2,
\nonumber\\
N &=& \sqrt{\frac{r}{32 m^3}} \; \mathrm{e}^\frac{r}{4m} \partial_u,
\end{eqnarray}
where for $r > 2m$, $u, v$ are
\begin{eqnarray}
u &=& \left( \frac{r}{2m} - 1 \right)^\frac12 \; \mathrm{e}^\frac{r}{4m} \sinh\frac{t}{4m},
\nonumber\\
v &=& \left( \frac{r}{2m} - 1 \right)^\frac12 \; \mathrm{e}^\frac{r}{4m} \cosh\frac{t}{4m},
\end{eqnarray}
and we find $N^t=A^{-1/2}\cosh\frac{t}{4m}$, so
 the value of the energy~(\ref{Energy}) is
\begin{equation}
E_{\rm KS} = r \left( 1 - \sqrt{1 - \frac{2m}{r}} \cosh\frac{t}{4m} \right), \label{EKSout}
\end{equation}
whereas for $r < 2m$, $u, v$ are
\begin{eqnarray}
u &=& \left( 1 - \frac{r}{2m} \right)^\frac12 \; \mathrm{e}^\frac{r}{4m} \cosh\frac{t}{4m},
\nonumber\\
v &=& \left( 1 - \frac{r}{2m} \right)^\frac12 \; \mathrm{e}^\frac{r}{4m} \sinh\frac{t}{4m},
\end{eqnarray}
then $N^t=(-A)^{-1/2}\cosh\frac{t}{4m}$ and the value of the energy~(\ref{Energy}) is
\begin{equation}
E_{\rm KS} = r \left( 1 + \sqrt{\frac{2m}{r} - 1} \; \cosh\frac{t}{4m} \right). \label{EKSin}
\end{equation}
The expressions for the two regions can be combined to give $N^t=|A|^{-1/2}\cosh\frac{t}{4m}$ and
\begin{equation}
E_{\rm KS} = r \left( 1 -{\rm sgn}(A) \sqrt{|A|} \; \cosh\frac{t}{4m} \right).\label{EKSboth}
\end{equation}
These KS results can be compared with those obtained by others using the Brown-York quasi-local energy expression. They agree with those in~\cite{Yu:2008ij} (aside from an essential missing absolute value sign) and also (at $t = 0$) with those in~\cite{Lundgren:2006fu}.

Whereas our first results (\ref{E-max},\ref{ETmax}) applied only to the region $r\ge 2m$ where there are  static observers,  it should be noted that all of our freely falling, EF, PG, and KS outcomes are well defined, smooth functions of $r$ through the horizon.


\subsection{Observer Adapted Coordinates}
In this section for any coordinate system we take the unit normal of the constant coordinate time hypersurface as the natural choice of the displacement vector. Now using a similar analysis we can find the natural coordinate system for any observer, which we call the observer adapted coordinates. For any $N = N^t \partial_t + N^r \partial_r$ we can find its orthogonal hypersurface. By choosing
\begin{eqnarray}
\vartheta^0 = a_0 dt + a_1 dr, \qquad \vartheta^1 = b_0 dt + b_1 dr,
\end{eqnarray}
and from $\vartheta^0(N) = 1, \vartheta^1(N) = 0$ we get
\begin{equation}
a_0 = \left( 1 \!-\! \frac{2m}{r} \right) N^t, \; a_1 = - \frac{N^r}{1 \!-\! \frac{2m}{r}}, \; b_0 =- N^r, \; b_1 =  N^t.
\end{equation}
Assume
$\vartheta^0 = F(t,r) du$, where $F^{-1}$ is an integrating factor, then
\begin{equation}
dt = \frac{F du}{\left( 1 - \frac{2m}{r} \right) N^t} + \left( 1 - \frac{2m}{r} \right)^{-2} \frac{N^r}{N^t} dr,
\end{equation}
and
\begin{equation}
ds^2 = - F^2 du^2 + 
\frac{1}{\left(1-\frac{2m}{r}\right)^2}\left[\frac{1}{N^t} dr- \frac{N^r}{N^t} F du   \right]^2 + r^2 d\Omega_2^2. \label{ds2AC}
\end{equation}
Now for any $N$ we can find the associated observer adapted coordinates in which the foliation is along the coordinate time, and $N$ is orthogonal to the constant time hypersurface. As an example let us consider the ingoing geodesic observers:
\begin{eqnarray}
N^t &=& \frac{1}{1 - \frac{2m}{r}} \frac{\sqrt{1 - \frac{2m}{a}}}{\sqrt{1 - v^2_0}},
\nonumber\\
N^r &=& -\left[ \frac{1 - \frac{2m}{a}}{1 - v_0^2} - \left( 1 - \frac{2m}{r} \right) \right]^\frac12,
\nonumber\\
F &=& 1.
\end{eqnarray}
Then
\begin{eqnarray}
ds^2 &=& - du^2 + \Biggl[\left( \frac{1 - v_0^2}{1 - \frac{2m}{a}} \right)^\frac12 dr
\nonumber\\
&& + \left( 1 - \frac{1 \!-\! \frac{2m}{r}}{1 \!-\! \frac{2m}{a}} (1 \!-\! v_0^2)\right)^{\frac12}du \Biggr]^2 + r^2 d\Omega_2^2.
\end{eqnarray}
For $v_0 = \sqrt{2m/a}$, which is equivalent to freely falling from infinity, this becomes the Painlev\'e-Gullstrand representation.

There are several different expressions which can give $m$ as the result of the quasi-local energy for the Schwarzschild spacetime~\cite{Bergqvist:1992a, Blau:2007wj}, so the coordinate system that gives this value should be of interest. By assigning a value $m$, from our energy expression~(\ref{Energy}) we find
\begin{equation} N^t = (1 - m/r)(1 - 2m/r)^{-1}, \qquad N^r = -m/r, \end{equation}
 and then calculating the integrating factor we get $F = 1 - m/r$. From~(\ref{ds2AC}) we find the form of the metric:
\begin{eqnarray}
ds^2 &=& - (1 - m/r)^2 du^2
\nonumber\\
&& + \left[(1 - m/r)^{-1} dr+ \frac{m}{r} du  \right]^2 + r^2 d\Omega_2^2.
\end{eqnarray}
This is the same result as that found in~\cite{Blau:2007wj} when the authors were searching for a foliation such that they can identify the Brown-York and the Misner-Sharp energy.

\subsection{The Meaning of Energy Extremization }
As an alternative to extremizing the energy, using~(\ref{genMetric}) we can instead require that the dynamic and reference frames completely match on the 2-surface:
\begin{eqnarray}
N &=& N^u \partial_u + N^v \partial_v = e_0 = \partial_T,
\nonumber\\
\vartheta^0 &=& a_1 du = \bar\vartheta^0 = \varphi^* dT,
\nonumber\\
\vartheta^1 &=& a_2 du + a_3 dv = \bar\vartheta^1 = \varphi^* dR, \label{isoframe}
\end{eqnarray}
here $\varphi^*$ means the pullback of the embedding map. For this kind of {\em isometric} matching requirement we find the conditions
\begin{eqnarray} \label{solT2R2n2}
&& T_u = a_1, \qquad T_v = 0,
\nonumber\\
&& R_u = a_2, \qquad R_v = a_3,
\nonumber\\
&& N^u = a_1^{-1}, \qquad N^v = - \frac{a_2}{a_1 a_3}.
\end{eqnarray}
From $-(\vartheta^0)^2 + (\vartheta^1)^2 + r^2 d\Omega_2^2 = ds^2$ we then get
\begin{eqnarray} \label{sola3}
a_1 &=& (A^{-1} r_v^2 - A t_v^2)^{-1/2} (t_u r_v - t_v r_u),
\nonumber\\
a_2 &=& (A^{-1} r_u r_v - A t_u t_v) (A^{-1} r_v^2 - A t_v^2)^{-1/2},
\nonumber\\
a_3 &=& (A^{-1} r_v^2 - A t_v^2)^{1/2}.
\end{eqnarray}
and
\begin{equation}
g^{uu} = - \frac1{a_1^2}, \quad g^{uv} = \frac{a_2}{a_1^2 a_3}, \quad g^{vv} = \frac{a_1^2 - a_2^2}{a_1^2 a_3^2}.
\end{equation}
Inserting these expressions along with the results~(\ref{solT2R2n2}, \ref{sola3}) into the energy expression~(\ref{expE}) gives the isometric matching energy:
\begin{equation}
E_{\rm iso} = r \left[ 1 - r_v ( A^{-1} r_v^2 - A t_v^2 )^{-1/2} \right].
\end{equation}
We now note that this energy value, with further use of~(\ref{solT2R2n2}, \ref{sola3}) can be rewritten as
\begin{eqnarray}
E_{\rm iso} &=& r \left[ 1 - \left( \frac{t_u}{a_1} - \frac{a_2 t_v}{a_1 a_3} \right) A \right]
\nonumber\\
&=& r \left[ 1 - (N^u t_u + N^v t_v) A \right]
\nonumber\\
&=& r (1 - N^t A) \equiv E_{\rm ex},
\end{eqnarray}
i.e., it turns out to have {\em exactly the same value} as the energy~(\ref{Energy}) found from the extremization program.

Now we can understand what the energy-extremization program has done for us: for any unit timelike displacement vector $N$ the program gives
us a 4D isometry on the two-sphere boundary between the dynamical and reference space,
and $N$ is the unit normal of the constant coordinate time hypersurface,
which is very much like the timelike Killing vector of the Minkowski reference being orthogonal to the constant $T$ hypersurface.

\section{Dynamic spacetime: FLRW cosmology}
\subsection{Energy-Extremization Program}

With the reference determined following the energy-extremization
program described in the previous section and in~\cite{Chen:2009zd},
%
%
we now apply the procedure to the dynamic homogeneous isotropic Friedmann-Lema\^{i}tre-Robertson-Walker (FLRW) spacetime. The FLRW metric is taken in the spherically symmetric form,
\begin{equation}
ds^2 = - dt^2 + \tilde{A}^2 dr^2 + a^2 (t)\,r^2\, d\Omega^2, \label{FLRW}
\end{equation}
where $\tilde{A} = {a(t)}/{\sqrt{1 - k r^2}}$ and $d\Omega^2 = d\theta^2 + \sin^2\theta d\phi^2$. Consider a more general version of the FLRW metric obtained by a coordinate transformation $t = t(u,v), r = r(u,v)$ then
\begin{eqnarray}
ds^2 &=& - (t_u^2 - \tilde{A}^2 r_u^2)\, du^2 + 2(\tilde{A}^2 r_u r_v - t_u t_v)\, du dv 
\nonumber\\
&& + (\tilde{A}^2 r_v^2 - t_v^2)\, dv^2 + a^2 r^2 d\Omega^2. \label{genmetric_flrw}
\end{eqnarray}
Note that we assume that the orientation is preserved under coordinate transformations so that the Jacobian is positive, i.e., $t_u r_v - t_v r_u > 0$. Choose Minkowski spacetime as the reference
\begin{equation}
d\bar{s}^2 = - dT^2 + dR^2 + R^2 d\Theta^2 + R^2 \sin^2\Theta d\Phi^2.
\end{equation}
We assume, in view of the spherical symmetry, $T = T(u,v), R = R(u,v), \Theta = \theta, \Phi = \phi$. This embeds a neighborhood of the boundary two-sphere $S$ at $(t_0, r_0)$ into Minkowksi space.  With the restriction  $R_0 := R(t_0, r_0) = a(t_0) r_0 = a_0 r_0$ the embedding is isometric on the boundary two-sphere. Assume~(\ref{0=DN}),
so that the second term of the quasi-local expression~(\ref{expB})
vanishes. For any given future timelike vector~(\ref{disp N})
from the expressions~(\ref{EN}, \ref{expB}) we find the quasi-local energy for a sphere of radius $r$ at time $t$ to be
\begin{eqnarray}
E(r, t)=\frac{\tilde{r}}{2} (N^u B + N^v C) \sqrt{-\tilde{\alpha}}, \label{Eexp_flrw}
\end{eqnarray}
where
\begin{eqnarray}
B &=& X T_u + g^{uv} (R_u - 2 \tilde{r}_u) + g^{vv} (R_v - 2 \tilde{r}_v),
\nonumber\\
C &=& X T_v + g^{uu} (2 \tilde{r}_u - R_u) + g^{uv}(2 \tilde{r}_v - R_v),
\nonumber\\
X &=& (T_u R_v - T_v R_u)^{-1},
\nonumber\\
\tilde{\alpha} &=& g_{uu} g_{vv} - g^2_{uv} = - \tilde A^2 (t_u r_v - t_v r_u)^2,
\nonumber\\
\tilde{r} &:=& a r.
\end{eqnarray}
Extremizing the energy with respect to the embedding variables again 
leads to the conditions~(\ref{ETu}--\ref{ERv}).
The conditions~(\ref{ETu}, \ref{ETv}) are equivalent (since we do not want both $T_u$ and $T_v$ to vanish), so we only have three independent restrictions. Using the relations
\begin{equation}
g^{uu} = \tilde{\alpha}^{-1} g_{vv}, \quad g^{uv} = - \tilde{\alpha}^{-1} g_{uv}, \quad g^{vv} = \tilde{\alpha}^{-1} g_{uu},
\end{equation}
conditions~(\ref{ETu}--\ref{ERv}) become
\begin{eqnarray}
N^u R_u + N^v R_v = N^R &=& 0,\quad\ \label{ETuv_flrw}
\\
X^2 T_v (N^u T_u + N^v T_v) - \tilde{\alpha}^{-1} (g_{uv} N^u + g_{vv} N^v) &=& 0,\quad\ \label{ERu2_flrw}
\\
X^2 T_u (N^u T_u + N^v T_v) - \tilde{\alpha}^{-1} (g_{uu} N^u + g_{uv} N^v) &=& 0.\quad\ \, \label{ERv2_flrw}
\end{eqnarray}
From $(\ref{ERv2_flrw}) \times R_v - (\ref{ERu2_flrw}) \times R_u$ we get
\begin{eqnarray}
X (N^u T_u + N^v T_v) + \tilde{\alpha}^{-1} \Bigl[ (g_{uv} N^u + g_{vv} N^v) R_u
\nonumber\\
- (g_{uu} N^u + g_{uv} N^v) R_v \Bigr] = 0. \label{ERuv_flrw}
\end{eqnarray}
Using~(\ref{ETuv_flrw}) we find
\begin{eqnarray}
R_u &=& - \frac{N^v}{N_u} R_v,
\nonumber \\
N^T &=& N^u T_u + N^v T_v = \frac{N^u}{X R_v}, \label{NTRu_flrw}
\end{eqnarray}
so the condition~(\ref{ERuv_flrw}) becomes
\begin{equation}
\frac{N^u}{R_v} - \tilde{\alpha}^{-1} \frac{R_v}{N_u} g(N,N) = 0 \quad \Rightarrow \quad R_v^2 = \frac{\tilde{\alpha} (N^u)^2}{g(N,N)}.
\end{equation}
We require $N^T > 0$, $N^u > 0$, and the Jacobians are positive. From~(\ref{NTRu_flrw}) we know $R_v$ should be positive, so
\begin{equation}
R_v = \sqrt{\frac{\tilde{\alpha}}{g(N,N)}} N^u, \quad R_u = - \sqrt{\frac{\tilde{\alpha}}{g(N,N)}} N^v. \label{RuRv_flrw}
\end{equation}
Now we can calculate the quasi-local energy. Using~(\ref{NTRu_flrw}, \ref{RuRv_flrw}) we find
\begin{eqnarray}
&& N^u B + N^v C
\nonumber\\
&=& \frac{N^u}{R_v} + \tilde{\alpha}^{-1} \Bigl[ \sqrt{\tilde{\alpha} g(N,N)} + 2 (g_{uv} N^u \tilde{r}_u
\nonumber\\
&& - g_{uu} N^u \tilde{r}_v + g_{vv} N^v \tilde{r}_u - g_{uv} N^v \tilde{r}_v) \Bigr]
\nonumber\\
&=& 2 \sqrt{\frac{g(N,N)}{\tilde{\alpha}}} \left( 1 \!-\! \sqrt{1 - k r^2} N^t \!-\! \frac{a \dot{a} r N^r}{\sqrt{1 \!-\! k r^2}} \right),
\end{eqnarray} 
where we have used the metric and
\begin{eqnarray}
\tilde{r}_u &=& a_u r + a r_u = \dot{a} t_u r + a r_u,
\nonumber\\
\tilde{r}_v &=& a_v r + a r_v = \dot{a} t_v r + a r_v, \label{tilder_flrw}
\end{eqnarray}
where $\dot{a} = {da}/{dt}$. Choose $N$ to be unit timelike on the two-sphere~(\ref{unitN}), then we get the quasi-local energy for a sphere of radius $r$ at time $t$ for any given future timelike displacement vector $N$:
\begin{equation}
E_{\rm ext}(N) = a r \left( 1 - \sqrt{1 - k r^2} N^t - \frac{a \dot{a} r}{\sqrt{1 - k r^2}} N^r \right). \label{energy_flrw}
\end{equation}
This result is independent of the coordinate choice and the embedding variables $T_u$ and $T_v$. Impose the normalization condition on the displacement vector in the reference spacetime $\bar{g}(N,N) = -1$, then we can uniquely determine the reference. Since $N^R = 0$ this condition means that $N$ is the timelike Killing vector field in the Minkowski reference, i.e., $N = \partial_T$. Using this condition and (\ref{ERu2_flrw}, \ref{ERv2_flrw}, \ref{RuRv_flrw}) to solve for $T_u$ and $T_v$ we find
\begin{eqnarray}
T_u &=& t_u N^t - \tilde{A}^2 r_u N^r,
\nonumber\\
T_v &=& t_v N^t - \tilde{A}^2 r_v N^r.
\end{eqnarray}
Now we can calculate the quasi-local energy of the FLRW spacetime for any physical observer. One obvious choice is $N = \partial_t$, i.e., the co-moving geodesic observer. From~(\ref{energy_flrw}) we get
\begin{equation}
E_{\rm ext}(\partial_t) = a r \left( 1 - \sqrt{1 - k r^2} \right)=\frac{akr^3}{1+\sqrt{1-kr^2}}. \label{negaE_flrw}
\end{equation}
This is the same as the result found from the Brown-York quasi-local energy expression~\cite{Afshar:2009pi} and also in earlier work by some of the present authors~\cite{Chen:2007gx}. One can see this energy value is positive, zero, or negative when $k=1, 0, -1$, respectively. It is noteworthy that the expression~(\ref{negaE_flrw}) {\em does not} give zero energy for the Milne universe, i.e., $k=-1$, and $a=t$, which is in fact diffeomorphic to Minkowski spacetime.

Furthermore, we can vary the energy with respect to the displacement vector. Since the displacement vector is unit timelike, $g(N,N) = - (N^t)^2 + \tilde{A}^2 (N^r)^2 = - 1$, we take
\begin{equation}
N^t = \cosh z, \quad \tilde{A} N^r = \sinh z,
\end{equation}
then we find
\begin{eqnarray}
\frac{\partial E_{\rm ext}}{\partial z} = 0 &\Rightarrow& \cosh z = \frac{\sqrt{1 - k r^2}}{\sqrt{1 - k r^2 - \dot{a}^2 r^2}},
\nonumber\\
&\Rightarrow& N = \frac{\sqrt{1 - k r^2}}{\sqrt{1 - k r^2 - \dot{a}^2 r^2}} \partial_t
\nonumber\\
&& \qquad - \frac{\dot{a} r}{a} \frac{\sqrt{1 - k r^2}}{\sqrt{1 - k r^2 - \dot{a}^2 r^2}} \partial_r, \label{dualmean_flrw}
\end{eqnarray}
and the extreme energy value is indeed maximum:
\begin{eqnarray}
E_{\rm max} &=& a r \left( 1 - \sqrt{1 - k r^2 - \dot{a}^2 r^2} \right)
\nonumber\\
&=& \frac{a r^3 (k + \dot{a}^2)}{1 + \sqrt{1 - k r^2 - \dot{a}^2 r^2}}, \label{Emax_flrw}
\\
\frac{\partial^2 E}{\partial z^2}\Bigr|_{\frac{\partial E}{\partial z}=0} &=& - a r \sqrt{1 - k r^2 - \dot{a}^2 r^2} \leq 0. \nonumber
\end{eqnarray}
One can see that this quasi-local energy is nonnegative using the Friedman equation $k + \dot{a}^2 = \frac{8 \pi}{3} \rho a^2$ (where $\rho$ is the energy density):
\begin{equation}
E_{\rm max} = \frac{\frac{8 \pi}{3} \rho a^3 r^3}{1 + \sqrt{1 - \frac{8 \pi}{3} \rho a^2 r^2}} \geq 0, \label{Epos_flrw}
\end{equation}
and it {\em vanishes} for the Milne universe. The corresponding displacement vector~(\ref{dualmean_flrw}) is a bit complicated, but it is actually just the dual mean curvature vector. The dual mean curvature vector $V_\bot$ is defined as
\begin{equation}
V_\bot = k_1 e_0 - k_0 e_1, \label{Dualmean_flrw}
\end{equation}
where $k_0$ and $k_1$ are the trace of the extrinsic curvature of the two-sphere boundary $S$ with respect to $e_0$ and $e_1$ respectively. We should also mention here the mean curvature vector $V$ is defined as
\begin{equation}
V = k_1 e_1 - k_0 e_0. \label{mean_flrw}
\end{equation}
These two vectors are independent of the choice of normal frames for $S$. They depend only on $S$ and constitute a set of natural normal vectors for $S$~\cite{Szabados:1994gw}.

Now let the displacement vector be 
the timelike Killing vector in the Minkowski reference, and then extremize the energy. From the preceding we expect that the result should be (\ref{dualmean_flrw}, \ref{Emax_flrw}). If this is indeed so our program would be even more satisfactory.
Assuming $N^T=\partial_T$ and $-1=g(N,N)$ once again leads to~(\ref{NuNv},\ref{X2}); then~(\ref{Eexp_flrw}) becomes
\begin{eqnarray}
E &=& \frac{\tilde{r}}{2} X \Bigl[ 1 - \tilde{\alpha}^{-1} X^{-2} + 2 \tilde{\alpha}^{-1} (g_{uv} \tilde{r}_u R_v - g_{uu} \tilde{r}_v R_u
\nonumber\\
&& - g_{vv} \tilde{r}_u R_u + g_{uv} \tilde{r}_v R_u) \Bigr] \sqrt{-\tilde{\alpha}}.
\end{eqnarray}
Now there are only two variables in our energy expression. We extremize the energy with respect to these two variables:
\begin{eqnarray}
\frac{\partial E}{\partial R_u} = 0 &\Rightarrow& (g_{vv} R_u - g_{uv} R_v) (1 + \tilde{\alpha}^{-1} X^{-2})
\nonumber\\
&& + 2 R_v (\tilde{r}_u R_v - \tilde{r}_v R_u) = 0, \label{ERuA_flrw}
\\
\frac{\partial E}{\partial R_v} = 0 &\Rightarrow& (g_{uu} R_v - g_{uv} R_u) (1 + \tilde{\alpha}^{-1} X^{-2})
\nonumber\\
&& - 2 R_u (\tilde{r}_u R_v - \tilde{r}_v R_u) = 0. \label{ERvA_flrw} 
\end{eqnarray}
Moreover, $(\ref{ERuA_flrw}) \times R_u + (\ref{ERvA_flrw}) \times R_v$ gives
\begin{eqnarray}
0 &=& (1 + \tilde{\alpha}^{-1} X^{-2}) (g_{uu} R_v^2 - 2 g_{uv} R_u R_v + g_{vv} R_u^2)
\nonumber\\
&=& - X^{-2} (1 + \tilde{\alpha}^{-1} X^{-2})
\nonumber\\
\Rightarrow \quad X &=& \sqrt{-\tilde{\alpha}^{-1}}.
\end{eqnarray}
Now~(\ref{ERuA_flrw}, \ref{ERvA_flrw}) tell us
\begin{equation}
\tilde{r}_u R_v - \tilde{r}_v R_u = 0, \label{rR_flrw}
\end{equation}
and together with~(\ref{X2}) we get
\begin{eqnarray}
R_u^2 &=& \frac{\tilde{r}_u^2}{g_{uu} \tilde{r}_v^2 - 2 g_{uv} \tilde{r}_u \tilde{r}_v + g_{vv} \tilde{r}_u^2}
\nonumber\\
&=& \frac{\tilde{r}_u^2}{1 - k r^2 - \dot{a}^2 r^2},
\nonumber\\
R_v^2 &=& \frac{\tilde{r}_v^2}{g_{uu} \tilde{r}_v^2 - 2 g_{uv} \tilde{r}_u \tilde{r}_v + g_{vv} \tilde{r}_u^2}
\nonumber\\
&=& \frac{\tilde{r}_v^2}{1 - k r^2 - \dot{a}^2 r^2},
\end{eqnarray}
where (\ref{tilder_flrw}) was used. Pick the plus sign, then we get the unique energy produced by this program:
\begin{eqnarray}
N &=& \frac{\sqrt{1 - k r^2}}{\sqrt{1 - k r^2 - \dot{a}^2 r^2}} \partial_t - \frac{\dot{a} r}{a} \frac{\sqrt{1 - k r^2}}{\sqrt{1 - k r^2 - \dot{a}^2 r^2}} \partial_r,
\nonumber\\
E(\partial_T)_{\rm ext} &=& a r \left( 1 - \sqrt{1 - k r^2 - \dot{a}^2 r^2} \right).
\end{eqnarray}
This result agrees with $E_{\rm max}$~(\ref{Emax_flrw}, \ref{Epos_flrw}).

\subsection{The Meaning of Energy Extremization and Its Outcomes}

Instead of extremizing the energy, one can require that the dynamic frame associated with the metric~(\ref{genmetric_flrw}) exactly matches the reference frame on the boundary: 
\begin{eqnarray}
N &=& N^u \partial_u + N^v \partial_v = e_0 = \partial_T,
\nonumber\\
\vartheta^0 &=& \tilde{a}_1 du = \bar\vartheta^0 = \varphi^* dT,
\nonumber\\
\vartheta^1 &=& \tilde{a}_2 du + \tilde{a}_3 dv = \bar\vartheta^1 = \varphi^* dR, \label{matching_flrw}
\end{eqnarray}
where $\varphi^*$ means pullback. From this kind of matching we get
\begin{eqnarray}
&& T_u = \tilde{a}_1, \quad T_v = 0, \quad R_u = \tilde{a}_2, \quad R_v = \tilde{a}_3,
\nonumber\\
&& N^u = \tilde{a}_1^{-1}, \qquad N^v = - \frac{\tilde{a}_2}{\tilde{a}_1 \tilde{a}_3}. \label{solT2R2n2_flrw}
\end{eqnarray} From $- (\vartheta^0)^2 + (\vartheta^1)^2 + r^2 d\Omega_2^2 = ds^2$, and picking the plus sign we get
\begin{eqnarray}
\tilde{a}_1 &=& \tilde{A} (\tilde{A}^2 r_v^2 - t_v^2)^{-1/2} (t_u r_v - t_v r_u),
\nonumber\\
\tilde{a}_2 &=& (\tilde{A}^2 r_u r_v - t_u t_v)(\tilde{A}^2 r_v^2 - t_v^2)^{-1/2},
\nonumber\\
\tilde{a}_3 &=& (\tilde{A}^2 r_v^2 - t_v^2)^{1/2}. \label{sola3_flrw}
\end{eqnarray}
Putting (\ref{solT2R2n2_flrw}, \ref{sola3_flrw}) into~(\ref{Eexp_flrw}) then gives
\begin{equation}
E_{\rm iso} = a r \left[ 1 - (\tilde{A}^2 r_v^2 - t_v^2)^{-1/2} (a r_v + \dot{a} r t_v) \right],
\end{equation}
On the other hand, transforming the value we found above in (\ref{energy_flrw}) yields
\begin{eqnarray}
E_{\rm ext}(\partial_T)&\equiv& a r \left[ 1 - \sqrt{1 - k r^2} N^t - \frac{a \dot{a} r}{\sqrt{1 - k r^2}} N^r \right]
\nonumber\\
&\equiv& a r \Bigl[ 1 - \sqrt{1 - k r^2} (N^u t_u + N^v t_v)
\nonumber\\
&& - \frac{a \dot{a} r}{\sqrt{1 - k r^2}} (N^u r_u + N^v r_v) \Bigr]
\nonumber\\
&\equiv& a r \left[ 1 - (\tilde{A}^2 r_v^2 - t_v^2)^{-1/2} (a r_v + \dot{a} r t_v) \right]\nonumber \\
&\equiv& E_{\rm iso}.
\end{eqnarray}
So for any unit timelike displacement vector $N$ this program gives us a 4D isometry at the two-sphere boundary between the dynamic geometry and the reference, and $N$ is the unit normal of the constant coordinate time hypersurface---which is similar to the timelike Killing vector of the Minkowski reference being orthogonal to the constant $T$ hypersurface.

From (\ref{Dualmean_flrw}, \ref{matching_flrw}) for this case we find
\begin{equation}
k_0 = 0, \qquad k_1 = 1.
\end{equation}
And from $N = \partial_T = \bar e_0$ we also know
\begin{equation}
\bar k_0 = 0, \qquad \bar k_1 = 1,
\end{equation}
where bar represents the reference quantities. Physically this means that the expansions of the two-sphere boundary $S$ and its image $\varphi(S)$ are the same, since the mean curvature vectors (aka expansion vectors) are the same.

Many investigators believe that quasi-local energy should be positive
(e.g.,~\cite{Liu:2003bx, Liu:2004dc}), 
 but one can see that for $k=-1$ the quasi-local energy (\ref{negaE_flrw}) is negative. On the other hand, the quasi-local value~(\ref{Epos_flrw}) is always positive. Thus the quasi-local energy value depends on the observer, but there is a set of observers in FLRW spacetime who would measure the maximum energy---which is nonnegative. Furthermore, if we wisely choose the reference for an observer, it should be possible to always get nonnegative energy values, since the physically meaningful energy is measured relative to the ground state.
 Although positivity is a nice property for gravitational energies, negative energies can also be reasonable, when due consideration is given to the dynamics, the different choices of observer and the ground state (for some discussion of this topic see Ref.~\cite{Nester:2008xd}). 

\section{Conclusion}
The covariant Hamiltonian quasi-local energy expression has several virtues, however it also suffers from two ambiguities: which displacement vector and which reference. We propose embedding the two-sphere boundary and its neighborhood in the dynamic spacetime into the Minkowski reference, with the 2-geometry being embedded isometrically, and then extremizing the energy to fix the embedding and thereby the reference. In this paper we use the Schwarzschild spacetime and FLRW cosmology to test this idea and obtained encouraging results.

For each future timelike displacement vector the program produces a uniquely corresponding energy, so that we can discuss the energy measured by different observers. For example, a radial geodesic observer in the Schwarzschild spacetime can measure the energy at any distance from the singularity---even inside the black hole. The energy measured can be positive, zero or even negative. We found that the negative result is related to the negative scalar curvature of the 3-hypersurface. By imposing some further conditions on the reference we
can
get a positive energy.
This program can also be applied to the Reissner-Norstr\"{o}m spacetime or other static spherically symmetric spacetimes.

For the dynamic FLRW cosmology, the sign of the energy of the co-moving observer is
$k$, the sign of the spatial curvature,
 which is negative for the open universe.
When
 we vary the energy with respect to the observers we find that the maximum is nonnegative.
 The physically meaningful energy is defined
 relative to a ground state. So in this sense, although positivity is a nice property for gravitational energies, negative energies
can also be reasonable, considering the dynamics, 
the different observers and the choice of the ground state. Moreover, we find that this program is actually equivalent to isometrically matching the 4D geometry at the two-sphere boundary and making the displacement vector orthogonal to the spacelike constant coordinate time hypersurface, just like the timelike Killing vector of the Minkowski reference. This matching has
a clear
geometrical meaning for FLRW cosmology: it matches the expansion of the two-sphere boundary
$S$ and that of its reference image $\varphi(S)$.

\section*{Acknowledgement}
This work was supported by the National Science Council of the R.O.C. under the grants NSC-99-2112-M-008-004 (JMN) and NSC 99-2112-M-008-005-MY3 (CMC) and in part by the National Center of Theoretical Sciences (NCTS).




\begin{references} 

\bibitem{tra}
  A.~Trautman,
  in {\it Gravitation: an Introduction to current research},
  ed. L.~Witten (Wiley, New York, 1958).

\bibitem{Papapetrou:1948jw}
  A.~Papapetrou,
  ``Einstein's Theory Of Gravitation And Flat Space,''
  Proc.\ Roy.\ Irish Acad.\ (Sect.\ A) {\bf 52A}, 11 (1948).

\bibitem{Bergmann:1953jz}
  P.~G.~Bergmann and R.~Thomson,
  ``Spin And Angular Momentum In General Relativity,''
  Phys.\ Rev.\  {\bf 89}, 400 (1953).

\bibitem{Mol58}
  C.~M{\o}ller,
  ``On the Localization of the Energy of a Physical System in the General Theory of Relativity,''
  Ann.~Phys.~{\bf 4}, 347 (1958).

\bibitem{Goldberg:1958zz}
  J.~N.~Goldberg,
  ``Conservation Laws in General Relativity,''
  Phys.\ Rev.\  {\bf 111}, 315 (1958).

\bibitem{lan}
  L.~D.~Landau and E.~M.~Lifshitz,
  {\it The Classical Theory of Fields} 2nd ed.
  (Addison-Wesley, Reading, MA, 1973).

\bibitem{wein}
  S.~ Weinberg,
  {\it Gravitation and Cosmology},
  (Wiley, New York, 1972).


\bibitem{MTW}
  C.~W.~Misner, K.~Thorne and J.~A.~Wheeler,
  {\it Gravitation},
  (Freeman, San Francisco, 1973).

\bibitem{Penrose:1982wp}
  R.~Penrose,
  ``Quasilocal mass and angular momentum in general relativity,''
  Proc.\ Roy.\ Soc.\ Lond.\  A {\bf 381}, 53 (1982).

\bibitem{Szabados:2004vb}
  L.~B.~Szabados,
  ``Quasi-local energy-momentum and angular momentum in GR: A review article,''
  Living Rev.\ Rel.\  {\bf 12}, 4 (2009).

\bibitem{Hawking:1968qt}
  S.~Hawking,
  ``Gravitational radiation in an expanding universe,''
  J.\ Math.\ Phys.\  {\bf 9}, 598 (1968).



\bibitem{Katz:1985}
  J.~Katz,
  ``A note on Komar's anomalous factor,''
  Class.\ Quant.\ Grav.\  {\bf 2}, 423 (1985).

\bibitem{Katz:1990}
  J.~Katz and A.~Ori,
  ``Localisation of field energy,''
  Class.\ Quant.\ Grav.\  {\bf 7}, 787 (1990).

\bibitem{Katz:2006uw}
  J.~Katz, D.~Lynden-Bell and J.~Bicak,
  ``Gravitational energy in stationary spacetimes,''
  Class.\ Quant.\ Grav.\  {\bf 23}, 7111 (2006)
  [arXiv:gr-qc/0610052].

\bibitem{Jezierski:1990vu}
  J.~Jezierski and J.~Kijowski,
  ``The localization of energy in gauge field theories and in linear gravitation,''
  Gen.\ Rel.\ Grav.\  {\bf 22}, 1283 (1990).

\bibitem{Dougan:1991zz}
  A.~J.~Dougan and L.~J.~Mason,
  ``Quasilocal mass constructions with positive energy,''
  Phys.\ Rev.\ Lett.\  {\bf 67}, 2119 (1991).

\bibitem{Bergqvist:1992b}
  G.~Bergqvist,
  ``Positivity and definitions of mass (general relativity),''
  Class.\ Quant.\ Grav.\  {\bf 9}, 1917 (1992).

\bibitem{Nester:1994zn}
  J.~M.~Nester and R.~S.~Tung,
  ``A quadratic spinor Lagrangian for general relativity,''
  Gen.\ Rel.\ Grav.\  {\bf 27}, 115 (1995)
  [arXiv:gr-qc/9407004].

\bibitem{Tung:1995cj}
  R.~S.~Tung and T.~Jacobson,
  ``Spinor one forms as gravitational potentials,''
  Class.\ Quant.\ Grav.\  {\bf 12}, L51 (1995)
  [arXiv:gr-qc/9502037].

\bibitem{Robinson:1996}
  D.~C.~Robinson,
  ``Spinor-valued forms and a variational principle for Einstein's vacuum equations,''
  Class.\ Quant.\ Grav.\  {\bf 13}, 307 (1996).

\bibitem{Hayward:1993ph}
  S.~A.~Hayward,
  ``Quasilocal gravitational energy,''
  Phys.\ Rev.\  D {\bf 49}, 831 (1994)
  [arXiv:gr-qc/9303030].

\bibitem{Brown:1992br}
  J.~D.~Brown and J.~W.~York,
  ``Quasilocal energy and conserved charges derived from the gravitational action,''
  Phys.\ Rev.\  D {\bf 47}, 1407 (1993)
  [arXiv:gr-qc/9209012].

\bibitem{Lau:1993yv}
  S.~Lau,
  ``Canonical variables and quasilocal energy in general relativity,''
  Class.\ Quant.\ Grav.\  {\bf 10}, 2379 (1993)
  [arXiv:gr-qc/9307026].

\bibitem{Bergqvist:1992a}
  G.~Bergqvist,
  ``Quasilocal mass for event horizons,''
  Class.\ Quant.\ Grav.\  {\bf 9}, 1753 (1992).

\bibitem{Chen:1994qg}
  C.~M.~Chen, J.~M.~Nester and R.~S.~Tung,
  ``Quasilocal energy momentum for gravity theories,''
  Phys.\ Lett.\  A {\bf 203}, 5 (1995)
  [arXiv:gr-qc/9411048].

\bibitem{Chen:1998aw}
  C.~M.~Chen and J.~M.~Nester,
  ``Quasilocal quantities for GR and other gravity theories,''
  Class.\ Quant.\ Grav.\  {\bf 16}, 1279 (1999)
  [arXiv:gr-qc/9809020].

\bibitem{Chang:1998wj}
  C.~C.~Chang, J.~M.~Nester and C.~M.~Chen,
  ``Pseudotensors and quasilocal gravitational energy-momentum,''
  Phys.\ Rev.\ Lett.\  {\bf 83}, 1897 (1999)
  [arXiv:gr-qc/9809040].

\bibitem{Chen:2000xw}
  C.~M.~Chen and J.~M.~Nester,
  ``A symplectic Hamiltonian derivation of quasilocal energy-momentum for GR,''
  Grav.\ Cosmol.\  {\bf 6}, 257 (2000)
  [arXiv:gr-qc/0001088].

\bibitem{Chen:2005hwa}
  C.~M.~Chen, J.~M.~Nester and R.~S.~Tung,
  ``The Hamiltonian boundary term and quasi-local energy flux,''
  Phys.\ Rev.\  D {\bf 72}, 104020 (2005)
  [arXiv:gr-qc/0508026].

\bibitem{Schoen:1979zz}
  R.~Schoen and S.~T.~Yau,
  ``Positivity of the total mass of a general space-time,''
  Phys.\ Rev.\ Lett.\  {\bf 43}, 1457 (1979).

\bibitem{Witten:1981mf}
  E.~Witten,
  ``A simple proof of the positive energy theorem,''
  Commun.\ Math.\ Phys.\  {\bf 80}, 381 (1981).

\bibitem{Liuthes}
J.~L.~Liu,
 ``On quasi-local energy and the choice of reference'',
 MSc.\ Thesis, National Central University, 2007;
%
 J.~L.~Liu, C.~M.~Chen and J.~M.~Nester,
Class.\ Quantum Grav. {\bf28}, 195019 (2011) [arXiv:1105.0502v2].


\bibitem{Chen:2009zd}
  C.~M.~Chen, J.~L.~Liu, J.~M.~Nester and M.~F.~Wu,
  ``Optimal Choices of Reference for Quasi-local Energy,''
  Phys.\ Lett. A \textbf{374}, 3599 (2010) [arXiv:0909.2754 [gr-qc]].



\bibitem{Booth:1998eh}
  I.~S.~Booth and R.~B.~Mann,
  ``Moving observers, non-orthogonal boundaries, and quasilocal energy,''
  Phys.\ Rev.\  D {\bf 59}, 064021 (1999)
  [arXiv:gr-qc/9810009].

\bibitem{Lundgren:2006fu}
  A.~P.~Lundgren, B.~S.~Schmekel and J.~W.~.~York,
  ``Self-renormalization of the classical quasilocal energy,''
  Phys.\ Rev.\  D {\bf 75}, 084026 (2007)
  [arXiv:gr-qc/0610088].

\bibitem{Yu:2008ij}
  P.~P.~Yu and R.~R.~Caldwell,
  ``Observer dependence of the quasi-local energy and momentum in Schwarzschild space-time,''
  Gen.\ Rel.\ Grav.\  {\bf 41}, 559 (2009)
  [arXiv:0801.3683 [gr-qc]].

\bibitem{Blau:2007wj}
  M.~Blau and B.~Rollier,
  ``Brown-York energy and radial geodesics,''
  Class.\ Quant.\ Grav.\  {\bf 25}, 105004 (2008)
  [arXiv:0708.0321 [gr-qc]].

\bibitem{Sharif:2008gq}
  M.~Sharif and M.~J.~Amir,
  ``Energy-Momentum of the Friedmann Models in General Relativity and Teleparallel Theory of Gravity,''
  Canadian J. Phys. {\bf 86}, 1297 (2008)
  [arXiv:0809.1529 [gr-qc]].


\bibitem{Afshar:2009pi}
  M.~M.~Afshar,
  ``Quasilocal Energy in FRW Cosmology,''
  Class.\ Quant.\ Grav.\  {\bf 26}, 225005 (2009)
  [arXiv:0903.3982 [gr-qc]].

\bibitem{Nester:2008xd}
  J.~M.~Nester, L.~L.~So and T.~Vargas,
  ``On the energy of homogeneous cosmologies,''
  Phys.\ Rev.\  D {\bf 78}, 044035 (2008)
  [arXiv:0803.0181 [astro-ph]].


\bibitem{Anco:2001gk}
  S.~C.~Anco and R.~S.~Tung,
  ``Symplectic structure of general relativity for spatially bounded spacetime regions. I: Boundary conditions,''
  J.\ Math.\ Phys.\  {\bf 43}, 5531 (2002)
  [arXiv:gr-qc/0109013].

\bibitem{Anco:2001gm}
  S.~C.~Anco and R.~S.~Tung,
  ``Symplectic structure of general relativity for spatially bounded spacetime regions. II: Properties and examples,''
  J.\ Math.\ Phys.\  {\bf 43}, 3984 (2002)
  [arXiv:gr-qc/0109014].

\bibitem{Liu:2003bx}
  C.~C.~Liu and S.~T.~Yau,
  ``New definition of quasilocal mass and its positivity,''
  Phys.\ Rev.\ Lett.\  {\bf 90}, 231102 (2003)
  [arXiv:gr-qc/0303019].

\bibitem{Wang:2008jy}
  M.~T.~Wang and S.~T.~Yau,
  ``Quasilocal mass in general relativity,''
  Phys.\ Rev.\ Lett.\  {\bf 102}, 021101 (2009)
  [arXiv:0804.1174 [gr-qc]].





\bibitem{Chen:2007gx}
  C.~M.~Chen, J.~L.~Liu and J.~M.~Nester,
  ``Quasi-local energy for cosmological models,''
  Mod.\ Phys.\ Lett.\  A {\bf 22}, 2039 (2007)
  [arXiv:0705.1080 [gr-qc]].


\bibitem{Szabados:1994gw}
  L.~B.~Szabados,
  ``Two-dimensional Sen connections and quasilocal energy momentum,''
  Class.\ Quant.\ Grav.\  {\bf 11}, 1847 (1994)
  [arXiv:gr-qc/9402005].


\bibitem{Liu:2004dc}
  C.~C.~Liu and S.~T.~Yau,
  ``Positivity of quasi-local mass II,''
  J. Amer.\ Math.\ Soc.\ {\bf19}, no.1, 181--204 (2006)
  [arXiv:math/0412292].




\end{references}
\end{document}